\documentclass[11pt,twoside,letterpaper]{article} 
\usepackage{times,fancyhdr}
\usepackage[dvips]{graphicx}
\usepackage{amsmath,epsfig,graphicx,amssymb}


\makeatletter 
\def\cleardoublepage{\clearpage\if@twoside \ifodd\c@page\else%
    \hbox{}%
    \thispagestyle{empty}%
    \newpage%
    \if@twocolumn\hbox{}\newpage\fi\fi\fi} 
\makeatother

\begin{document}
\title{
{\begin{flushleft}
\vskip 0.45in
{\normalsize\bfseries\textit{Chapter~1}}
\end{flushleft}
\vskip 0.45in
\bfseries\scshape Pseudo-critical behavior of spin-1/2 Ising diamond and tetrahedral chains}}
\author{\bfseries\itshape Jozef Stre\v{c}ka\thanks{E-mail address: jozef.strecka@upjs.sk}\\
Department of Theoretical Physics and Astrophysics \\ Faculty of Science of P.~J.~\v{S}af\'arik University \\ Ko\v{s}ice, Slovak Republic}
\date{}
\maketitle
\thispagestyle{empty}
\setcounter{page}{1}
\thispagestyle{fancy}
\fancyhead{}
\fancyhead[L]{In: Book Title \\ 
Editor: Editor Name, pp. {\thepage-\pageref{lastpage-01}}} 
\fancyhead[R]{ISBN 0000000000  \\
\copyright~2006 Nova Science Publishers, Inc.}
\fancyfoot{}
\renewcommand{\headrulewidth}{0pt}

\begin{abstract} 
A few paradigmatic one-dimensional lattice-statistical spin models have recently attracted a vigorous scientific interest owing to their peculiar thermodynamic behavior, which is highly reminiscent of a temperature-driven phase transition. The pseudo-transitions of one-dimensional lattice-statistical spin models differ from actual phase transitions in several important aspects: the first-order derivatives of the Gibbs free energy such as entropy or magnetization exhibit near a pseudo-transition an abrupt continuous change instead of a true discontinuity, whereas the second-order derivatives of the Gibbs free energy such as specific heat or susceptibility display near a pseudo-transition a vigorous finite peak instead of an actual power-law divergence. In the present chapter we will comprehensively examine a pseudo-critical behavior of the spin-1/2 Ising diamond and tetrahedral chains by a detailed examination of basic magnetothermodynamic quantities such as the entropy, specific heat and susceptibility. It will be demonstrated that density plots of these magnetothermodynamic quantities provide a useful tool for establishing a finite-temperature diagram, which clearly delimits boundaries between individual quasi-phases in spite of a lack of true spontaneous long-range order at any nonzero temperature. It is suggested that a substantial difference between the degeneracies of two ground states of the spin-1/2 Ising diamond and tetrahedral chains is an essential prerequisite for observation of a relevant pseudo-critical behavior in a close vicinity of their ground-state phase boundary.
\end{abstract}

\vspace{2in}

\noindent \textbf{PACS} 75.10.Hk, 75.30.Sg, 75.40.-s, 75.50.Cc. 
\vspace{.08in} \noindent \textbf{Keywords:} Ising model, pseudo-transition, diamond chain, tetrahedral chain, spin frustration.


\pagestyle{fancy}
\fancyhead{}
\fancyhead[EC]{Jozef Stre\v{c}ka}
\fancyhead[EL,OR]{\thepage}
\fancyhead[OC]{Pseudo-critical behavior of spin-1/2 Ising diamond and tetrahedral chains}
\fancyfoot{}
\renewcommand\headrulewidth{0.5pt} 

\section{Introduction}

The history of the Ising model dates back a century ago when Lenz introduced \cite{len20} and Ising exactly solved \cite{isi25} one-dimensional version of this paradigmatic lattice-statistical model, which was originally aimed at a theoretical description of a magnetic phase transition of ferromagnetic materials \cite{nis05}. However, the exact solution of the one-dimensional spin-1/2 Ising model disproved presence of a temperature-driven phase transition at any non-zero temperature and the same conclusion was initially reached also for its higher-dimensional versions \cite{isi25}. This faulty conjecture significantly suppressed an investigation of the Ising model until the famous Onsager's exact solution \cite{ons44} for the spin-1/2 Ising model on a rectangular lattice rigorously proved an intriguing temperature-driven phase transition elusive to Landau theory of phase transition. The Ising model nowadays represents one of the most studied lattice-statistical models in statistical mechanics with regard to its diverse applications in different areas of science a long ago surpassing ordinary physics branches (see Ref. \cite{str15} for a recent review). 

Generally, one-dimensional lattice-statistical models with short-range interactions and non-singular potential do not exhibit a phase transition at any non-zero temperature in agreement with the non-existence theorems proved by van Hove \cite{hov50}, Cuesta and S\'anchez \cite{cue04}. However, a recent investigation of a few prototypical one-dimensional lattice-statistical models has verified anomalous thermodynamical behavior, which is highly reminiscent of a temperature-driven phase transition \cite{sou17,roj19,kro19}. Among a few paradigmatic examples of the one-dimensional lattice-statistical models displaying this peculiar phenomenon one could mention a double-tetrahedral chain involving the localized Ising spins and mobile electrons \cite{gal15}, the spin-1/2 Ising-Heisenberg three-leg tube \cite{str16} and two-leg ladder \cite{roj16}, the spin-1/2 Ising-XYZ diamond chain \cite{car19} and the mixed spin-1/2 and spin-1 Ising-Heisenberg double-tetrahedral chain \cite{roj20}. It should be emphasized, moreover, that pseudo-transitions of the aforementioned class of one-dimensional lattice-statistical models differ from actual phase transitions in several important aspects. The first-order derivatives of the Gibbs free energy as for instance entropy or magnetization exhibit a very steep but still continuous change instead of a true discontinuity close to a pseudo-critical temperature, while the second-order derivatives of the Gibbs free energy as for instance specific heat or susceptibility display a sharp but still finite peak instead of a true power-law divergence in this region \cite{sou17,roj19,kro19}. On the other hand, the pseudo-transitions of one-dimensional lattice-statistical models resemble a true phase transition through a substantial rise of the correlation length spread over several orders of magnitudes (though it still remains finite) \cite{car19}, whereas sizable peaks of the second-order derivatives of the Gibbs free energy are governed sufficiently close (but not too close) to a pseudo-critical temperature by a power law characterized through a universal set of pseudo-critical exponents \cite{roj19}. 

In this chapter we will extend a set of one-dimensional lattice-statistical models displaying a remarkable pseudo-critical behavior by considering another two prototypical examples of  the spin chains, which could be referred to as the spin-1/2 Ising diamond and tetrahedral chains. The spin-1/2 Ising diamond chain in a magnetic field represents a special limiting case of the exactly solved spin-1/2 Ising-Heisenberg diamond chain \cite{can04,can06}, whose thermodynamic properties have not been comprehensively studied yet and the pseudo-critical behavior was only very briefly reported in our preliminary study \cite{str20}. On the other hand, the magnetic properties and pseudo-critical behavior of the spin-1/2 Ising tetrahedral chain as a special limiting case of the exactly solved spin-1/2 Ising-Heisenberg tetrahedral chain \cite{roj13,str14} remain completely unexplored yet.

\section{Spin-1/2 Ising diamond chain in a magnetic field}

\begin{figure}[t]
\begin{center}
\includegraphics[width=0.75\textwidth]{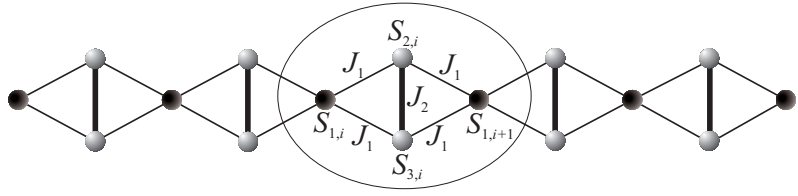}
\end{center}
\vspace{-0.6cm}
\caption{A schematic illustration of the spin-1/2 Ising diamond chain. An oval demarcates the $i$th diamond spin cluster defined through the Hamiltonian (\ref{hami}).}
\label{fig1}
\end{figure}

First, we will examine a pseudo-critical behavior of the symmetric spin-1/2 Ising diamond chain in a magnetic field diagrammatically depicted in Fig.~\ref{fig1}, which can be mathematically defined through the following Hamiltonian:
\begin{eqnarray}
{\cal H} = J_2 \sum_{i=1}^N S_{2,i} S_{3,i} + J_1 \sum_{i=1}^N (S_{2,i} + S_{3,i}) (S_{1,i} + S_{1,i+1}) - h \sum_{i=1}^N \sum_{j=1}^3 S_{j,i}.
\label{ham}
\end{eqnarray}
The Hamiltonian (\ref{ham}) involves two-valued Ising spin variables $S_{j,i} = \pm 1/2$, whereas the latter subscript $i$ determines unit cell of an underlying diamond chain and the former subscript $j$ a position within the $i$th unit cell. For easy notation, the Ising spins assigned to a monomeric (dimeric) site specified by the subscript $j=1$ ($j=2$ or $3$) will be hereafter referred to as the monomeric (dimeric) spins, respectively. In the spirit of this notation the coupling constant $J_1$ denotes the nearest-neighbor interaction between the monomeric and dimeric spins (i.e. the parameter $J_1$ is a monomer-dimer coupling constant), while the coupling constant $J_2$ stands for the nearest-neighbor interaction between the dimeric spins (i.e. the parameter $J_2$ is an intra-dimer coupling constant). The last term in the Hamiltonian (\ref{ham}) denotes the standard Zeeman's term closely connected with the magnetostatic energy of the Ising spins (and associated magnetic moments) in a presence of the external magnetic field $h$. For the sake of simplicity, the periodic boundary condition $S_{1,N+1} \equiv S_{1,1}$ is considered.

It should be mentioned that the spin-1/2 Ising diamond chain has been exactly solved in the past as a special limiting case of a more general spin-1/2 Ising-Heisenberg diamond chain \cite{can04,can06} by making use of a generalized decoration-iteration mapping transformation \cite{fis59,roj09,str10}, which establishes a rigorous mapping correspondence with the effective spin-1/2 Ising chain. In spite of this fact, the magnetic properties of the spin-1/2 Ising diamond chain were not comprehensively investigated yet \cite{can04,can06} and the relevant pseudo-critical behavior of the spin-1/2 Ising diamond chain was reported only very recently in a preliminary study \cite{str20}. In this chapter we will adapt at first a transfer-matrix approach due to Kramers and Wannier \cite{kra41} in order to derive an exact solution for the spin-1/2 Ising diamond chain in a magnetic field defined through the Hamiltonian (\ref{ham}) by more standard means. To this end, let us decompose the Hamiltonian (\ref{ham}) into a sum taken over the cluster Hamiltonians ${\cal H} = \sum_{i=1}^N {\cal H}_i$, whereas the $i$th cluster Hamiltonian 
${\cal H}_i$ involves all interaction terms pertinent to a single diamond spin cluster schematically demarcated in Fig.~\ref{fig1} by an oval:
\begin{eqnarray}
{\cal H}_i = J_2 S_{2,i} S_{3,i} + J_1 (S_{2,i} + S_{3,i}) (S_{1,i} + S_{1,i+1}) - h (S_{2,i} + S_{3,i}) - \frac{h}{2} (S_{1,i} + S_{1,i+1}).
\label{hami}
\end{eqnarray}
It is worthwhile to remark that the factor $1/2$ at the last term avoids a double counting of the Zeeman term pertinent to the monomeric spins, which is always symmetrically split into two neighboring diamond spin clusters. The partition function of the spin-1/2 Ising diamond chain in a magnetic field can be factorized into the following useful form:
\begin{eqnarray}
{\cal Z} = \sum_{\{S_{1,i} \}}\prod_{i = 1}^{N} \sum_{S_{2,i} = \pm \frac{1}{2}} \sum_{S_{3,i} = \pm \frac{1}{2}} \exp(-\beta {\cal H}_i),
\label{pf}
\end{eqnarray}
where $\beta=1/(k_{\rm B} T)$, $k_{\rm B}$ is Boltzmann's constant, $T$ is the absolute temperature and the symbol $\sum_{\{S_{1,i}\}}$ denotes a summation over all available spin configurations of a full set of the monomeric  spins. The structure of the relation (\ref{pf}) is compatible with the fact that one may perform summation over spin degrees of freedom of a couple of the dimeric spins $S_{2,i}$ and $S_{3,i}$ from the $i$th unit cell independently with respect to the other ones and before performing a summation over spin degrees of freedom of all monomeric spins $\{S_{1,i}\}$. Apparently, the summation over spin degrees of freedom of a couple of the dimeric spins $S_{2,i}$ and $S_{3,i}$ gives the effective Boltzmann's weight:
\begin{eqnarray}
\boldsymbol{T} (S_{1,i}, S_{1,i+1}) \!\!\!&=&\!\!\! \sum_{S_{2,i} = \pm \frac{1}{2}} \sum_{S_{3,i} = \pm \frac{1}{2}} \exp(-\beta {\cal H}_i) 
= 2 \exp \left[\frac{\beta h}{2} (S_{1,i} + S_{1,i+1}) \right] \nonumber \\
\!\!\!&\times&\!\!\!
\left\{\exp \left(\!-\frac{\beta J_2}{4} \right) \cosh \left[\beta J_1 (S_{1,i} + S_{1,i+1}) - \beta h \right] + \exp \left(\frac{\beta J_2}{4} \right) \right\}\!\!,
\label{tm}
\end{eqnarray}
which depends on two nearest-neighbor monomeric spins ($S_{1,i}$, $S_{1,i+1}$) and can be further identified as the transfer matrix with the following definition of the individual transfer-matrix elements:
\begin{eqnarray}
\boldsymbol{T}(S_{1,i}, S_{1,i+1}) = 
\left(\!
\begin{array}{cc} 
   \boldsymbol{T} \left(+\frac{1}{2},+\frac{1}{2}\right) & \boldsymbol{T} \left(+\frac{1}{2},-\frac{1}{2}\right)    \\
   \boldsymbol{T} \left(-\frac{1}{2},+\frac{1}{2}\right) & \boldsymbol{T} \left(-\frac{1}{2},-\frac{1}{2}\right)    \\
\end{array}
                 \! \right)  =
\left(\!
\begin{array}{cc} 
   V_1     &     V_3    \\
   V_3     &     V_2    \\
\end{array}
                 \! \right).
\label{tmixice}
\end{eqnarray}
For the sake of brevity, we have introduced in Eq. (\ref{tmixice}) the following notation for three different  transfer-matrix elements defined through the functions:
\begin{eqnarray}
V_1 \!\!\!&=&\!\!\! \boldsymbol{T}\left(+\frac{1}{2},+\frac{1}{2}\right) = 2 \exp \left(\frac{\beta h}{2}\right) \left[\exp \left(-\frac{\beta J_2}{4} \right) 
\cosh \left(\beta J_1 - \beta h \right) + \exp \left(\frac{\beta J_2}{4} \right) \right], \nonumber \\
V_2 \!\!\!&=&\!\!\! \boldsymbol{T}\left(-\frac{1}{2},-\frac{1}{2}\right) = 2 \exp \left(-\frac{\beta h}{2}\right) \left[\exp \left(-\frac{\beta J_2}{4} \right) 
\cosh \left(\beta J_1 + \beta h \right) + \exp \left(\frac{\beta J_2}{4} \right) \right], \nonumber \\
V_3 \!\!\!&=&\!\!\! \boldsymbol{T}\left(\pm\frac{1}{2},\mp\frac{1}{2}\right) = 2 \left[\exp \left(-\frac{\beta J_2}{4} \right) 
\cosh \left(\beta h \right) + \exp \left(\frac{\beta J_2}{4} \right) \right].
\label{tmixicev}
\end{eqnarray}
Now, one may proceed in Eq. (\ref{pf}) to subsequent summation over spin degrees of freedom of a full set of the monomeric spins $\{S_{1,i}\}$, which enables one to express the partition function of the spin-1/2 Ising diamond chain in a magnetic field in terms of two eigenvalues $\lambda_{\pm}$ of the transfer matrix (\ref{tmixice}) when using the standard transfer-matrix formalism \cite{kra41}: 
\begin{eqnarray}
{\cal Z}  = \sum_{\{S_{1,i} \}} \prod_{i=1}^N  \boldsymbol{T} (S_{1,i}, S_{1,i+1}) 
= \sum_{S_{1,1} = \pm \frac{1}{2}} \!\! \boldsymbol{T}^N(S_{1,1}, S_{1,1}) 
= \mbox{Tr} \, \boldsymbol{T}^N = \lambda_{+}^N + \lambda_{-}^N. 
\label{pfmixicr}
\end{eqnarray}
The eigenvalues $\lambda_{\pm}$ can be readily obtained by a direct diagonalization of the two-by-two transfer matrix (\ref{tmixice}):
\begin{eqnarray}
\lambda_{\pm} = \frac{1}{2} \left[V_1 + V_2 \pm \sqrt{(V_1 - V_2)^2 + 4 V_3^2} \right].
\label{pfmixev}
\end{eqnarray}
In the thermodynamic limit $N \to \infty$, the Gibbs free energy of the spin-1/2 Ising diamond chain in a magnetic field  can be expressed solely in terms of the larger transfer-matrix eigenvalue ($\lambda_{+}>\lambda_{-}$): 
\begin{eqnarray}
G  = - k_{\rm B} T \lim_{N \to \infty} \ln {\cal Z} = - N k_{\rm B} T \ln \lambda_{+}.  
\label{gmix}	
\end{eqnarray} 
All basic magnetothermodynamic quantities such as the entropy $S$, the specific heat $C$, the magnetization $M$, and the susceptibility $\chi$ can be now easily calculated with the help of standard relations of thermodynamics and statistical physics: 
\begin{eqnarray}
S = - \left(\frac{\partial G}{\partial T}\right)_h, \quad  
C = - T \left(\frac{\partial^2 G}{\partial T^2}\right)_h, \quad
M = - \left(\frac{\partial G}{\partial h}\right)_T, \quad
\chi = - \left(\frac{\partial^2 G}{\partial h}\right)_T.  
\label{scms}	
\end{eqnarray}

\begin{figure}[t]
\begin{center}
\includegraphics[width=0.6\textwidth]{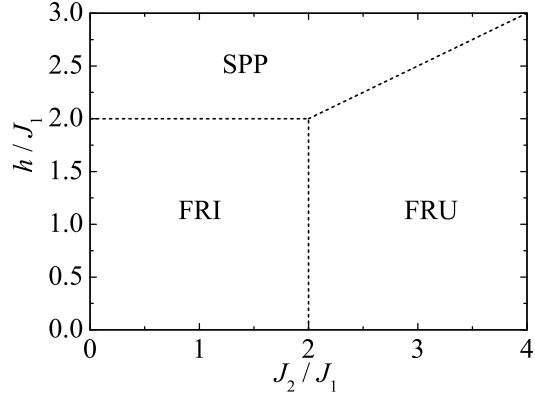}
\end{center}
\vspace{-1.0cm}
\caption{The ground-state phase diagram of the symmetric spin-1/2 Ising diamond chain in the $J_2/J_1 - h/J_1$ plane. The notation for individual ground states: the ferrimagnetic phase - FRI, the frustrated phase - FRU, the saturated paramagnetic phase - SPP.}
\label{gsdc}
\end{figure}

Before proceeding to a discussion of the most interesting results for a pseudo-critical behavior of the antiferromagnetic spin-1/2 Ising diamond chain ($J_1 > 0, J_2 > 0$) in a magnetic field let us make a few comments on the ground-state phase diagram, which is depicted in Fig. \ref{gsdc} in the $J_2/J_1 - h/J_1$ plane \cite{can04}. Note that all the interaction terms are normalized in Fig. \ref{gsdc} with respect to the monomer-dimer coupling constant $J_1$, which will from here onward serve as an energy unit. The ground-state phase diagram of the spin-1/2 Ising diamond chain in a magnetic field involves in total three different ground states, which can be classified according to the underlying spin arrangements as: \\
i. the ferrimagnetic phase 
\begin{eqnarray}
|\mbox{FRI} \rangle = \prod_{i=1}^N \left|-\frac12 \right\rangle_{\!\!1,i} \left|\frac12 \right\rangle_{\!\!2,i} \left|\frac12 \right\rangle_{\!\!3,i},  
\label{fri}	
\end{eqnarray}
ii. the frustrated phase 
\begin{eqnarray}
|\mbox{FRU} \rangle = \prod_{i=1}^N \left|\frac12 \right\rangle_{\!\!1,i} \left|\pm \frac12 \right\rangle_{\!\!2,i} \left|\mp \frac12 \right\rangle_{\!\!3,i}, 
\label{fru}	
\end{eqnarray}
iii. the saturated paramagnetic phase 
\begin{eqnarray}
|\mbox{SPP} \rangle = \prod_{i=1}^N \left|\frac12 \right\rangle_{\!\!1,i} \left|\frac12 \right\rangle_{\!\!2,i}\left|\frac12 \right\rangle_{\!\!3,i}. 
\label{spp}	
\end{eqnarray}
The ferrimagnetic phase represents the respective ground state of the spin-1/2 Ising diamond chain in the parameter region $J_2/J_1<2$ and $h/J_1<2$, the frustrated phase is the relevant ground state in the parameter space $J_2/J_1>2$ and $h/J_1< 1 + J_2/(2J_1)$ and finally, the saturated paramagnetic phase forms the respective ground state in the rest of the parameter space. It is noteworthy that the term 'frustrated' relates to a two-fold degeneracy of each couple of the dimeric spins, which are antiferromagnetically aligned with respect to each other according to Eq. (\ref{fru}). In the consequence of that, the frustrated phase (\ref{fru}) exhibits a macroscopic degeneracy reflected in the nonzero residual entropy $S = N k_{\rm B} \ln 2$.  

The spin-1/2 Ising diamond chain exhibits a pseudo-critical behavior just if a highly degenerate manifold of low-lying excited states pertinent to the frustrated phase (\ref{fru}) is formed above the non-degenerate ferrimagnetic ground state (\ref{fri}). This situation is ensured by a suitable choice of the Hamiltonian parameters $J_2/J_1 \lesssim 2$ and $h/J_1 < 2$, which drive the spin-1/2 Ising diamond chain sufficiently close to a ground-state phase boundary between the ferrimagnetic (\ref{fri}) and frustrated (\ref{fru}) phases. From the physical point of view, the pseudo-critical behavior closely relates to intense thermal excitations from the quasi-ferrimagnetic phase (qFRI) to the quasi-frustrated phase (qFRU), which originate from an energy closeness and high entropic difference of a non-degenerate ferrimagnetic ground state (\ref{fri}) and a macroscopically degenerate manifold of the first excited state inherent to the frustrated phase (\ref{fru}). The term 'quasi' is used, because the spin-1/2 Ising diamond chain cannot exhibit a true spontaneous long-range order at any finite temperature due to its one-dimensional character. In accordance with this picture, the spin-1/2 Ising diamond chain displays a pseudo-transition at the pseudo-critical temperature $k_{\rm B} T_p/J_1 = (2 - J_2/J_1)/\ln 4$, which can be derived from a direct comparison of the free energies of the quasi-ferrimagnetic and quasi-frustrated phases when simply ignoring changes of the enthalpy and entropy at sufficiently low temperatures (e.g. $k_{\rm B} T_p/J_1 \approx 0.036$ for $J_2/J_1 = 1.95$) \cite{gal15,str16}.

\begin{figure}[t]
\begin{center}
\includegraphics[width=0.51\textwidth]{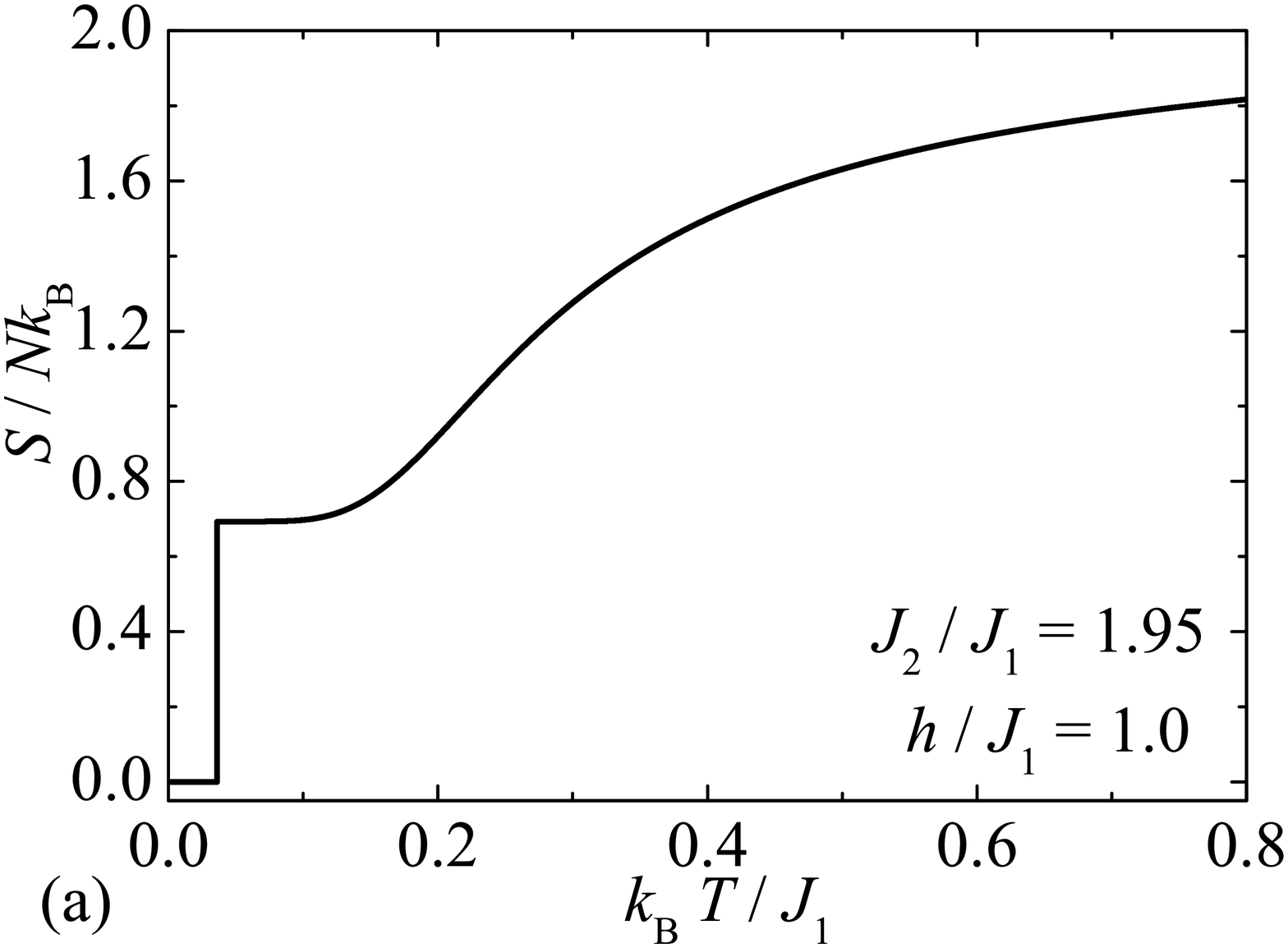}
\hspace*{-0.6cm}
\includegraphics[width=0.51\textwidth]{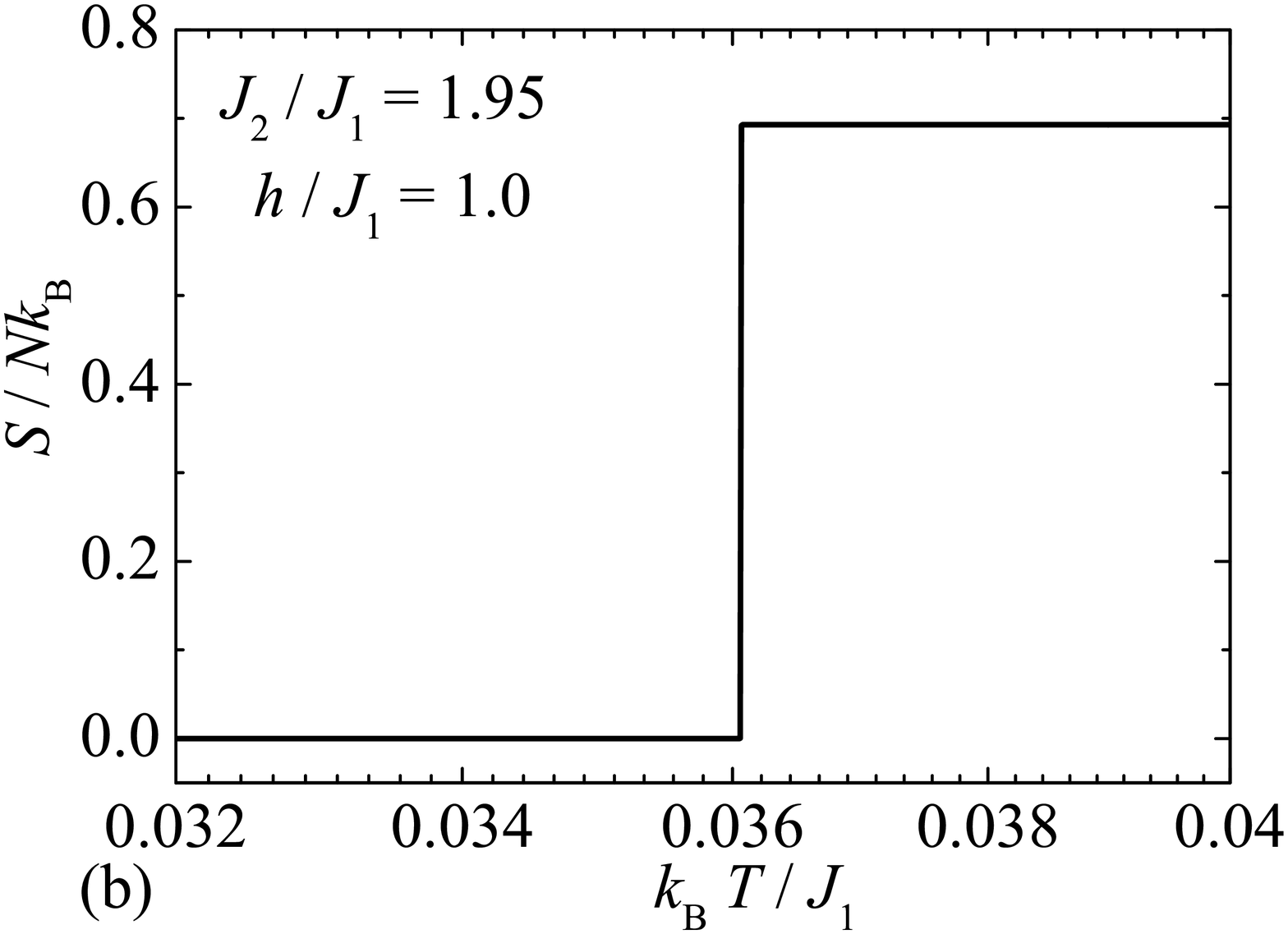}
\end{center}
\vspace{-0.7cm}
\caption{The temperature dependence of the entropy of the symmetric spin-1/2 Ising diamond chain for the particular value of the interaction ratio $J_2/J_1 = 1.95$ and the magnetic field
$h/J_1 = 1.0$. Panel (b) shows a detailed plot of the entropy in a vicinity of the pseudo-transition, where the entropy exhibits an abrupt but still continuous change quite reminiscent of a discontinuous jump.}
\label{dcen}
\end{figure}

\begin{figure}[t]
\begin{center}
\includegraphics[width=0.51\textwidth]{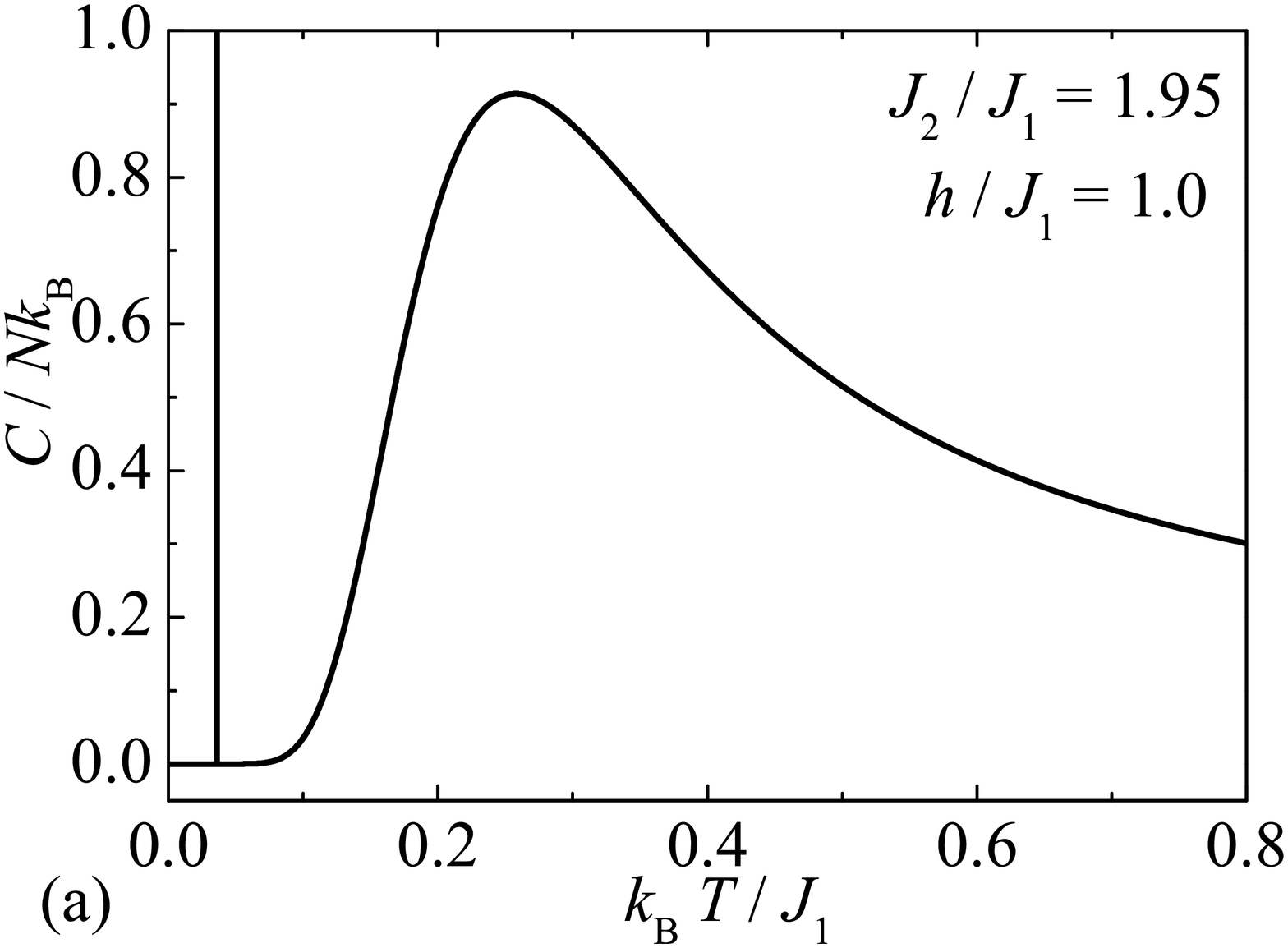}
\hspace*{-0.6cm}
\includegraphics[width=0.51\textwidth]{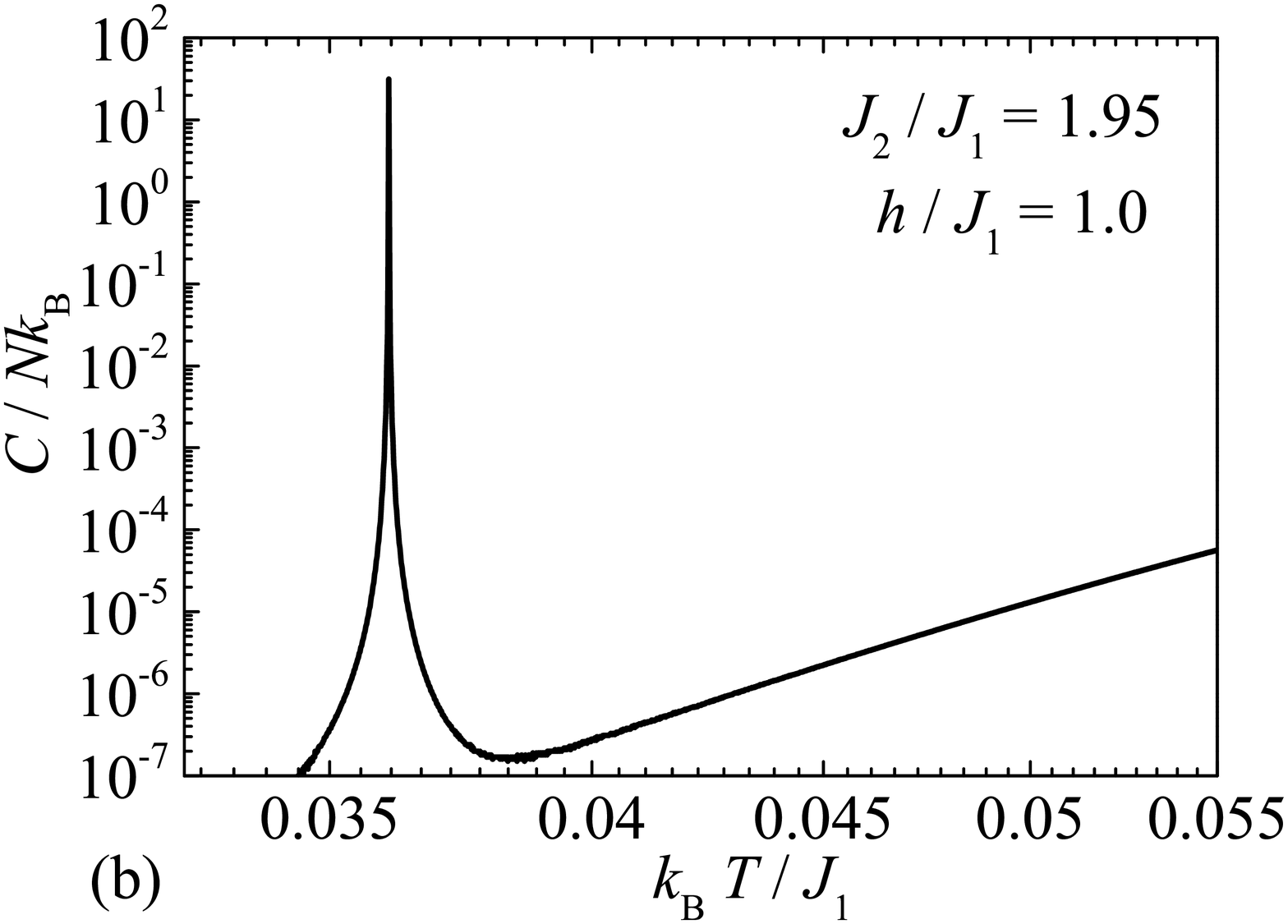}
\end{center}
\vspace{-0.7cm}
\caption{The temperature dependence of the specific heat of the symmetric spin-1/2 Ising diamond chain for the particular value of the interaction ratio $J_2/J_1 = 1.95$ and the magnetic field $h/J_1 = 1.0$. Panel (b) shows a detailed semi-logarithmic plot of the specific heat in a vicinity of the pseudo-transition, where the specific heat exhibit a sharp finite peak quite reminiscent of a power-law divergence.}
\label{dcsh}
\end{figure}

To gain a deeper insight into a pseudo-critical behavior of the spin-1/2 Ising diamond chain, temperature variations of the entropy and specific heat are plotted in Figs. \ref{dcen} and \ref{dcsh} for the suitable choice of the interaction ratio $J_2/J_1 = 1.95$ and the magnetic field $h/J_1 = 1.0$. It can be seen from  Fig. \ref{dcen} that an abrupt change in the respective temperature dependence of the entropy at the pseudo-critical temperature $k_{\rm B} T_p/J_1 \approx 0.036$ is followed by a quasi-plateau emergent around the value  $S \approx N k_{\rm B} \ln 2$, which is in agreement with the residual entropy (macroscopic degeneracy) of the frustrated phase. Note furthermore that an abrupt but still continuous change of the entropy is quite reminiscent of a discontinuous jump and it thus mimics a discontinuous temperature-driven phase transition. On the other hand, the specific heat exhibits a striking thermal dependence with a very sharp robust low-temperature peak, which is well separated from a round high-temperature maximum [see Fig. \ref{dcsh}(a)]. In this regard, the sharp sizable peak of the specific heat emergent at a pseudo-critical temperature mimics a power-law divergence accompanying a continuous temperature-driven phase transition. It should be pointed out, however, that the origin of this sharp sizable but still finite peak lies in intense thermal excitations from the quasi-ferrimagnetic phase to the quasi-frustrated phase rather than in a true continuous phase transition. 

\begin{figure}[t]
\begin{center}
\includegraphics[width=0.51\textwidth]{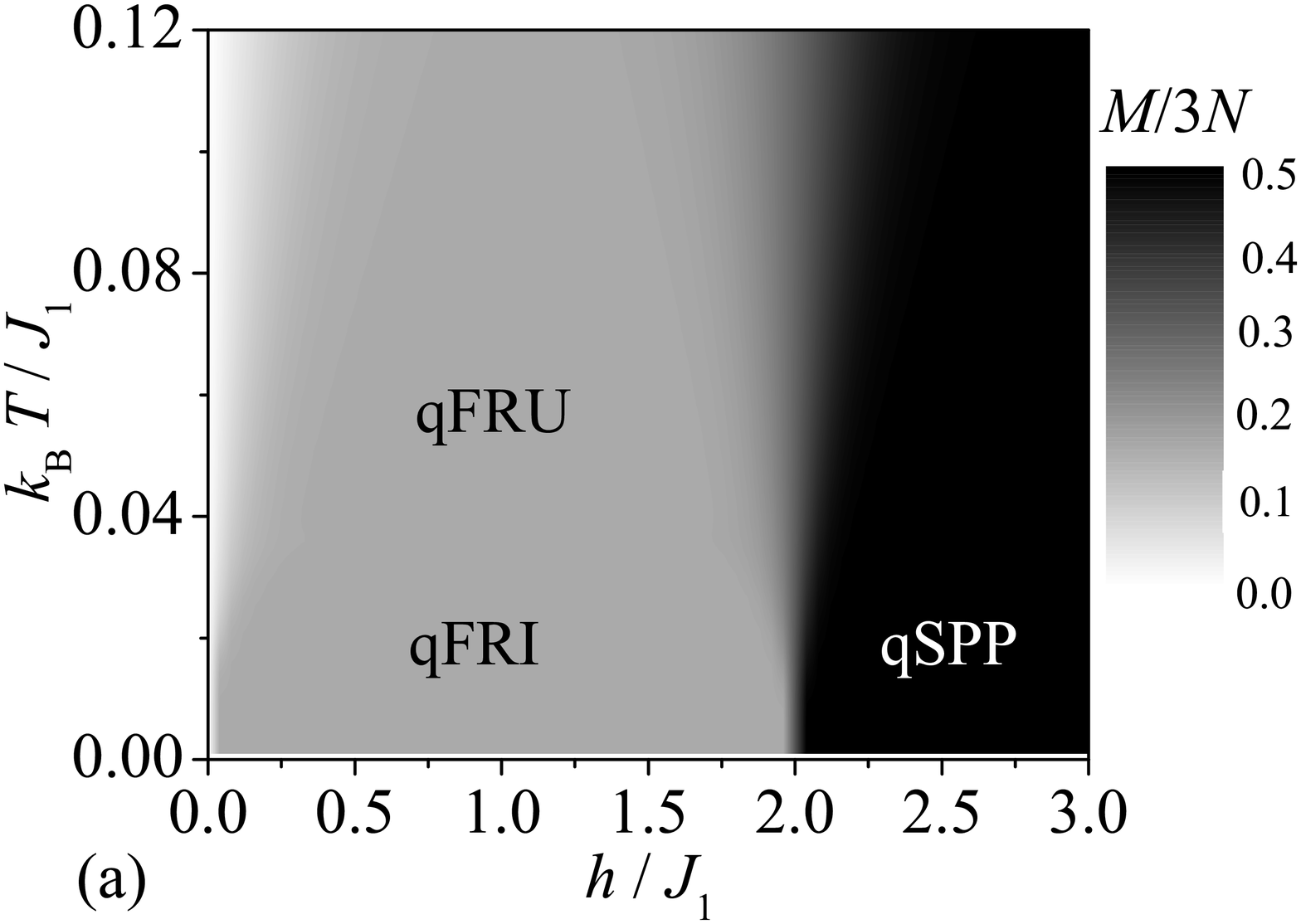}
\hspace*{-0.6cm}
\includegraphics[width=0.51\textwidth]{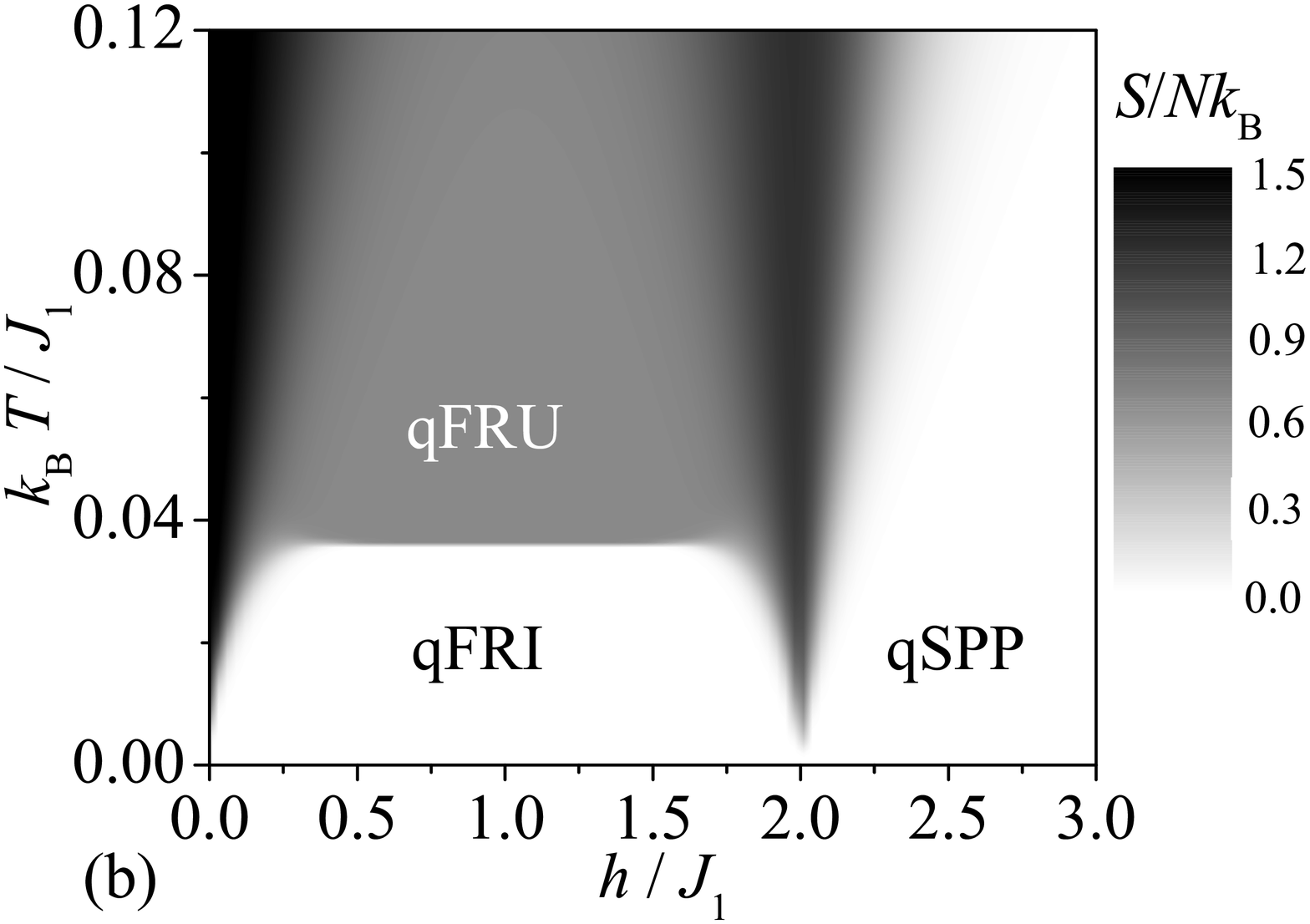}
\end{center}
\vspace{-0.7cm}
\caption{A density plot of the magnetization [panel (a)] and entropy [panel (b)] of the symmetric spin-1/2 Ising diamond chain in the field-temperature 
($h/J_1-k_{\rm B} T/J_1$) plane for the particular value of the interaction ratio $J_2/J_1 = 1.95$.}
\label{dc3dms}
\end{figure}

It is obvious from Figs. \ref{dcen} and \ref{dcsh} that the most pronounced signatures of a pseudo-transition of the spin-1/2 Ising diamond chain can be traced back from the relevant thermal variations of basic thermodynamic quantities such as the entropy and specific heat. Our further attention will be accordingly focused on a question whether or not a pseudo-transition of the spin-1/2 Ising diamond chain can be detected with the help of basic magnetic quantities such as the magnetization and susceptibility. For a comparison, the density plots of the magnetization and entropy of the spin-1/2 Ising diamond chain in the field-temperature plane are displayed in Fig. \ref{dc3dms} for one selected value of the interaction ratio $J_2/J_1 = 1.95$. While no traces of the pseudo-transition of the spin-1/2 Ising diamond chain cannot be found in the respective density plot of the total magnetization [Fig. \ref{dc3dms}(a)], the density plot of the entropy [Fig. \ref{dc3dms}(b)] affords a clear evidence of the pseudo-transition emergent at the pseudo-critical temperature $k_{\rm B} T_p/J_1 \approx 0.036$. In addition, it can be observed from Fig. \ref{dc3dms}(b) that the pseudo-transition of the spin-1/2 Ising diamond chain can be detected in the most authentic manner for the moderate values of the magnetic field $h/J_1 \in (0.5, 1.5)$, while the pseudo-transition apparently melts at lower ($h/J_1 \lesssim 0.5$) and higher ($h/J_1 \gtrsim 1.5$) when it is spread over a wider temperature interval.    

\begin{figure}[t]
\begin{center}
\includegraphics[width=0.51\textwidth]{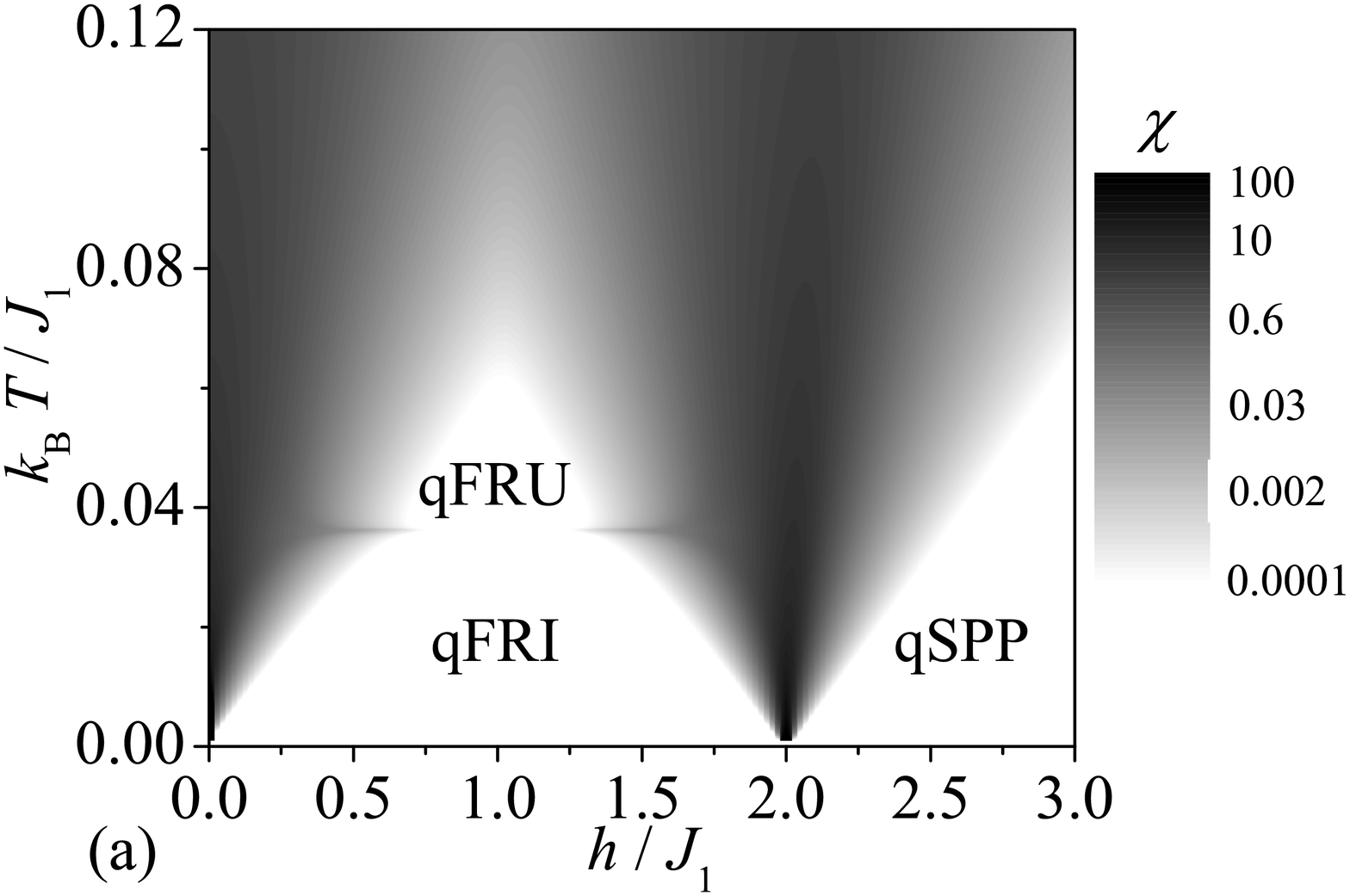}
\hspace*{-0.6cm}
\includegraphics[width=0.51\textwidth]{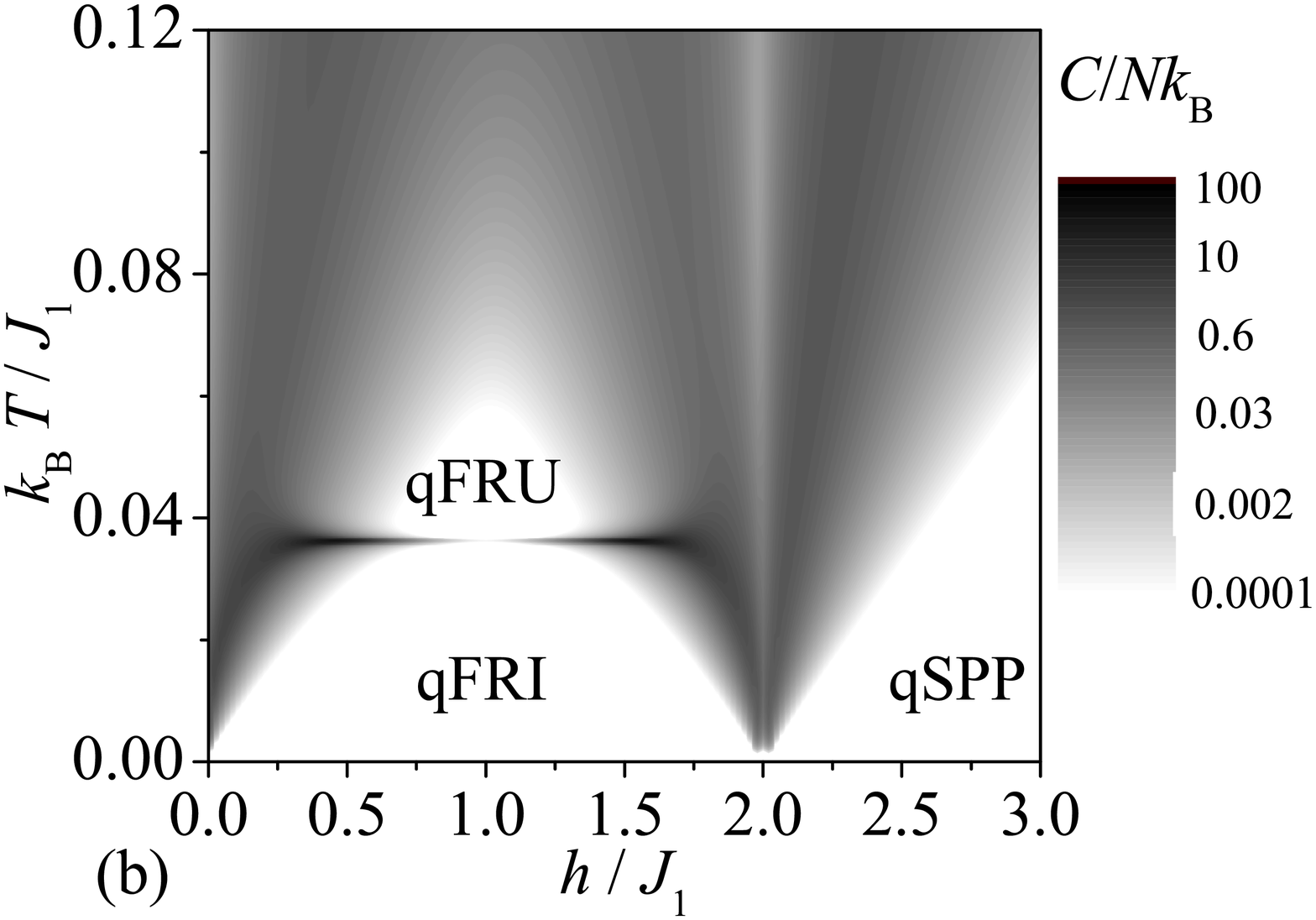}
\end{center}
\vspace{-0.7cm}
\caption{A density plot of the magnetic susceptibility [panel (a)] and specific heat [panel (b)] of the symmetric spin-1/2 Ising diamond chain in the field-temperature ($h/J_1-k_{\rm B} T/J_1$) plane for the particular value of the interaction ratio $J_2/J_1 = 1.95$.}
\label{dc3dss}
\end{figure}

Last but not least, the density plots of the susceptibility and specific heat of the spin-1/2 Ising diamond chain in the field-temperature plane are depicted in Fig. \ref{dc3dss} for the same value of the interaction ratio $J_2/J_1 = 1.95$. It can be observed from Fig. \ref{dc3dss} that the susceptibility and specific heat exhibit qualitatively the same feature: both these magnetothermodynamic quantities display a substantial rise spread over several orders of magnitude in a vicinity of the pseudo-critical temperature $k_{\rm B} T_p/J_1 \approx 0.036$. From this perspective, the susceptibility and specific heat represent the best suited quantities for a construction of the finite-temperature phase diagram of the spin-1/2 Ising diamond chain in a magnetic field, since they show the most distinct changes when temperature drives the investigated spin system across a pseudo-transition. In fact, the density plots of the susceptibility and specific heat shown Fig. \ref{dc3dss}(a) and (b) clearly allocate all individual quasi-phases. The quasi-ferrimagnetic (qFRI) phase can be located inside of a dome with relatively small values of the susceptibility and specific heat (white regions). This dome is bounded from the right by a quasi-critical fan emanating from the quasi-phase boundary with the quasi-saturated-paramagnetic (qSPP) phase centered at $h/J_1 = 2.0$ and from above by a curvilinear quasi-phase boundary with the quasi-frustrated (qFRU) phase.  

\section{Spin-1/2 Ising tetrahedral chain in a magnetic field}

\begin{figure}[t]
\begin{center}
\includegraphics[width=0.6\textwidth]{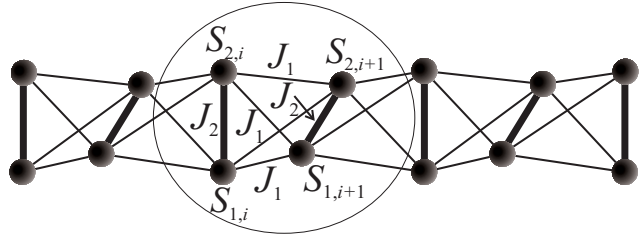}
\end{center}
\vspace{-0.6cm}
\caption{A schematic illustration of the spin-1/2 Ising tetrahedral chain composed of edge-sharing tetrahedra. An oval demarcates the $i$th tetrahedral spin cluster defined through the Hamiltonian 
(\ref{hamit}).}
\label{figt}
\end{figure}

In the following part, we will investigate a pseudo-critical behavior of the spin-1/2 Ising tetrahedral chain in a magnetic field diagrammatically depicted in Fig.~\ref{figt}, which can be mathematically defined through the Hamiltonian:
\begin{eqnarray}
\label{eq:Ham-orig}
{\cal H} = \sum_{i=1}^{N}\left[J_{2}S_{1,i}S_{2,i}+J_{1}(S_{1,i}+S_{2,i})(S_{1,i+1}+S_{2,i+1}) - h (S_{1,i}+S_{2,i}) \right],
\end{eqnarray}
where $S_{j,i} = \pm 1/2$  mark two Ising spin variables ($j=1,2$) assigned to the $i$th unit cell. The interaction term $J_2$ denotes the Ising interaction assigned to edges shared by the neighboring tetrahedra, the interaction parameter $J_{1}$ stands for the Ising interaction assigned to all other edges and $h$ is the standard Zeeman's term connected with the external magnetic field. For the sake of simplicity, the periodic boundary conditions $S_{j,N+1} \equiv S_{j,1} (j=1,2)$ are considered. It is worthwhile to remark that the spin-1/2 Ising tetrahedral chain represents a special case of the exactly solved spin-1/2 Ising-Heisenberg tetrahedral chain \cite{roj13,str14}, whose magnetic properties and pseudo-critical behavior have been also completely overlooked \cite{roj13,str14}. 

The Hamiltonian (\ref{eq:Ham-orig}) of the spin-1/2 Ising tetrahedral chain can be decomposed into a sum taken over the cluster Hamiltonians ${\cal H} = \sum_{i=1}^N {\cal H}_i$, whereas the $i$th cluster Hamiltonians ${\cal H}_i$ involves all interaction terms pertinent to a single tetrahedral spin cluster schematically demarcated in Fig.~\ref{figt} by an oval:
\begin{eqnarray}
{\cal H}_i \!\!\!&=&\!\!\! \frac{J_2}{2} \left(S_{1,i} S_{2,i} + S_{1,i+1} S_{2,i+1} \right) + J_1 (S_{1,i} + S_{2,i}) (S_{1,i+1} + S_{2,i+1}) \nonumber \\ 
\!\!\!&-&\!\!\! \frac{h}{2} (S_{1,i} + S_{2,i} + S_{1,i+1} + S_{2,i+1}).
\label{hamit}
\end{eqnarray}
Note that the factor $1/2$ at the first and third terms of the cluster Hamiltonian (\ref{hamit}) avoids a double counting of the the interaction term $J_1$ and Zeeman's term $h$, which are symmetrically split into two neighboring tetrahedral spin clusters. The partition function of the spin-1/2 Ising tetrahedral chain in a magnetic field can be subsequently factorized into the following useful form:
\begin{eqnarray}
{\cal Z} = \sum_{\{S_{1,i} \}} \sum_{\{S_{2,i} \}} \prod_{i = 1}^{N} \exp(-\beta {\cal H}_i) 
         = \sum_{\{S_{1,i} \}} \sum_{\{S_{2,i} \}} \prod_{i = 1}^{N} \boldsymbol{T} (S_{1,i}, S_{2,i}; S_{1,i+1}, S_{2,i+1}) 
				 = \mathrm{Tr} \, \boldsymbol{T}^N,
\label{pft}
\end{eqnarray}
where $\beta=1/(k_{\rm B} T)$, $k_{\rm B}$ is Boltzmann's constant, $T$ is the absolute temperature, while the symbols $\sum_{\{S_{1,i}\}}$ and $\sum_{\{S_{2,i}\}}$ denotes summations over all available spin configurations of the Ising spins. In the latter step we have implemented a consecutive summation over four available spin configurations of a couple of the Ising spins from the same unit cell within the standard transfer-matrix approach \cite{kra41}, which allows to express the partition function of the spin-1/2 Ising tetrahedral chain in a magnetic field in terms of the transfer matrix:
\begin{eqnarray}
\boldsymbol{T} (S_{1,i}, S_{2,i}; S_{1,i+1}, S_{2,i+1}) \!\!\!&=&\!\!\! \langle S_{1,i}, S_{2,i} | 
\exp(-\beta {\cal H}_i) | S_{1,i+1}, S_{2,i+1} \rangle \nonumber \\ 
\!\!\!&=&\!\!\! \left(\begin{array}{cccc}
xyz^2 & z & z &  xy^{-1} \\
z & x^{-1} & x^{-1} & z^{-1}\\
z & x^{-1} & x^{-1} & z^{-1}\\
xy^{-1} & z^{-1} & z^{-1} & xyz^{-2}
\end{array}\right)
\label{tmt}
\end{eqnarray}
with $x=\exp(-\beta J_2/4)$, $y=\exp(-\beta J_{1})$, and $z=\exp(\beta h/2)$. As usual, the largest eigenvalue of the transfer matrix (\ref{tmt}) determines in the thermodynamic limit $N \to \infty$ an exact result for the partition function, Gibbs free energy and overall thermodynamics. Apparently, one out of four eigenvalues of the transfer matrix (\ref{tmt}) equals zero ($\lambda_0 = 0$) since its second and third row are linearly dependent, while other three eigenvalues can be found by solving the characteristic equation:
\begin{equation}
\lambda^3 - a \lambda^2 + b \lambda + c =0
\label{eq:cubic-eq}
\end{equation}
with the coefficients
\begin{eqnarray}
a \!\!&=&\!\! 2 + 2 \exp\left(-\beta J_1 - \frac{\beta J_2}{2}\right) \cosh (\beta h), \nonumber \\
b \!\!&=&\!\! 4 [-1 + \exp\left(-\beta J_1\right)] \exp\left(-\frac{\beta J_2}{2}\right) \cosh (\beta h) 
       - 2 \exp\left(-\beta J_2\right) \sinh (2 \beta J_1), \nonumber \\
c \!\!&=&\!\! 4 \exp\left(-\beta J_2\right) [\sinh (2 \beta J_1) - \sinh (\beta J_1)].
\label{eq:abc}
\end{eqnarray}
The remaining three transfer-matrix eigenvalues then readily follow from the solution of the cubic equation (\ref{eq:cubic-eq}):
\begin{eqnarray}
\lambda_{j} =  \frac{a}{3} + 2 \mbox{sgn}(q) \sqrt{p} \cos \left[\frac{1}{3} \left(\phi + j 2 \pi \right)\right]\!, \,\,\, (j=1,2,3),
\label{rce}
\end{eqnarray}
whereas 
\begin{eqnarray}
p = \left( \frac{a}{3} \right)^2 - \frac{b}{3}, \qquad 
q = \left( \frac{a}{3} \right)^3 - \frac{ab}{6} - \frac{c}{2}, \qquad
\phi = \arctan \left(\frac{\sqrt{p^3 - q^2}}{q} \right).
\label{rcec}
\end{eqnarray}
In the thermodynamic limit $N \to \infty$, the Gibbs free energy of the spin-$1/2$ Ising tetrahedral chain is given by the largest eigenvalue $\lambda_{\rm max} = {\rm max} \{\lambda_1, \lambda_2, \lambda_3\}$ among the three roots (\ref{rce}) of the cubic equation (\ref{eq:cubic-eq}):
\begin{equation}
G = - k_{\rm B} T \lim_{N \to \infty} \ln {\cal Z} = -\frac{N J_{2}}{4}- N k_{\rm B} T \ln \lambda_{\rm max}.
\label{fe} 
\end{equation}
Other basic magnetothermodynamic quantities such as the entropy $S$, the specific heat $C$, the magnetization $M$, and the susceptibility $\chi$ can be subsequently calculated with the help of standard thermodynamical relations (\ref{scms}) listed in the previous part.

\begin{figure}[t]
\begin{center}
\includegraphics[width=0.6\textwidth]{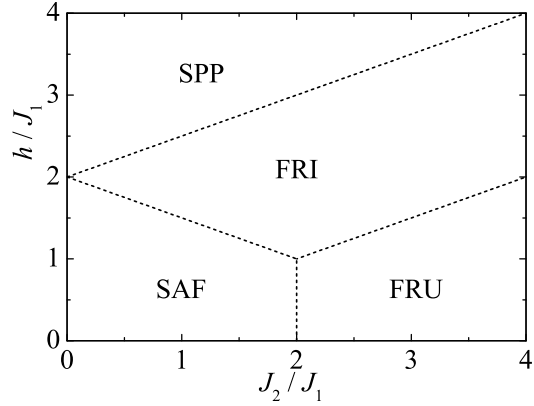}
\end{center}
\vspace{-1.0cm}
\caption{The ground-state phase diagram of the spin-1/2 Ising tetrahedral chain in the $J_2/J_1 - h/J_1$ plane. The notation for individual ground states: the superantiferromagnetic phase - SAF, the frustrated antiferromagnetic phase - FRU, the frustrated ferrimagnetic phase - FRI, the saturated paramagnetic phase - SPP.}
\label{gst}
\end{figure}

Now, let us make a few comments on the most interesting results obtained for a pseudo-critical behavior of the spin-1/2 Ising tetrahedral chain in a magnetic field. Before doing this,  let us at first describe the ground-state phase diagram of the spin-1/2 Ising tetrahedral chain, which is displayed in Fig. \ref{gst} in the $J_2/J_1 - h/J_1$ plane. It should be noted that all the interaction terms are normalized with respect to the coupling constant $J_1$, which will henceforth serve as an energy unit. The ground-state phase diagram of the spin-1/2 Ising tetrahedral chain in a magnetic field totally includes four different ground states, which can be classified according to the underlying spin arrangements as:\\ 
i. the superantiferromagnetic phase 
\begin{eqnarray}
|\mbox{SAF} \rangle = \prod_{i=1}^{N/2} \left|\frac12 \right\rangle_{\!\!1,2i-1} \left|\frac12 \right\rangle_{\!\!2,2i-1} 
                                        \left|-\frac12 \right\rangle_{\!\!1,2i}  \left|-\frac12 \right\rangle_{\!\!2,2i},  
\label{tsaf}	
\end{eqnarray}
ii. the frustrated antiferromagnetic phase 
\begin{eqnarray}
|\mbox{FRU} \rangle = \prod_{i=1}^N \left|\pm\frac12 \right\rangle_{\!\!1,i} \left|\mp \frac12 \right\rangle_{\!\!2,i}, 
\label{tfru}	
\end{eqnarray}
iii. the frustrated ferrimagnetic phase 
\begin{eqnarray}
|\mbox{FRI} \rangle = \prod_{i=1}^{N/2} \left|\frac12 \right\rangle_{\!\!1,2i-1} \left|\frac12 \right\rangle_{\!\!2,2i+1} 
                                        \left|\pm\frac12 \right\rangle_{\!\!1,2i}  \left|\mp\frac12 \right\rangle_{\!\!2,2i},  
\label{tfri}	
\end{eqnarray}
iv. the saturated paramagnetic phase 
\begin{eqnarray}
|\mbox{SPP} \rangle = \prod_{i=1}^N \left|\frac12 \right\rangle_{\!\!1,i} \left|\frac12 \right\rangle_{\!\!2,i}. 
\label{tspp}	
\end{eqnarray}
It is worth noticing that the other linearly independent eigenvectors corresponding to the superantiferromagnetic and the frustrated ferrimagnetic phase can be obtained from the eigenvectors (\ref{tsaf}) and (\ref{tfru}) upon interchange of the spin states on odd and even unit cells. The two-fold degenerate superantiferromagnetic (SAF) phase represents the respective ground state of the spin-1/2 Ising tetrahedral chain in the parameter region $J_2/J_1<2$ and $h/J_1<2 - J_2/(2J_1)$, the macroscopically degenerate frustrated antiferromagnetic (FRU) phase is the relevant ground state in the parameter space $J_2/J_1>2$ and $h/J_1< J_2/(2J_1)$, the non-degenerate saturated paramagnetic (SPP) phase is stable above the saturation field $h/J_1> 2 + J_2/(2J_1)$, and finally, the macroscopically degenerate frustrated ferrimagnetic (FRI) phase is the relevant ground state in the rest of the parameter region. It should be stressed that the term 'frustrated' relates to a two-fold degeneracy of antiferromagnetically aligned Ising spins, which are present according to Eqs. (\ref{tfru}) and (\ref{tfri}) within the frustrated antiferromagnetic or ferrimagnetic phase either on each or each second unit cell. Consequently, the residual entropy $S = N k_{\rm B} \ln 2$ of the frustrated antiferromagnetic phase (\ref{tfru}) is twice as large as the residual entropy $S = \frac{1}{2} N k_{\rm B} \ln 2$ of the frustrated ferrimagnetic phase (\ref{tfri}) with the alternating character of the unit cells (non-degenerate versus two-fold degenerate).  

It turns out that the spin-1/2 Ising tetrahedral chain shows a pseudo-critical behavior if a highly degenerate manifold of low-lying excited states pertinent to the frustrated antiferromagnetic phase (\ref{tfru}) exist above the two-fold degenerate superantiferromagnetic ground state (\ref{tsaf}). This situation is met whenever the interaction parameters $J_2/J_1 \lesssim 2$ and $h/J_1 < 1$ drive the spin-1/2 Ising tetrahedral chain close enough to a ground-state phase boundary between the superantiferromagnetic phase (\ref{tsaf}) and the frustrated antiferromagnetic phase (\ref{tfru}). The corresponding pseudo-critical behavior can be repeatedly attributed to intense thermal excitations, which originate from an energy closeness and high entropic difference of the superantiferromagnetic ground state (\ref{tsaf}) and a highly degenerate manifold of the first excited state inherent to the frustrated antiferromagnetic phase (\ref{tfru}). The spin-1/2 Ising tetrahedral chain accordingly exhibits a pseudo-transition at the pseudo-critical temperature $k_{\rm B} T_p/J_1 = (2 - J_2/J_1)/\ln 4$, which follows from a direct comparison of the free energies of the quasi-superantiferromagnetic and quasi-frustrated-antiferromagnetic phases when neglecting changes of the internal energy and entropy at sufficiently low temperatures (e.g. $k_{\rm B} T_p/J_1 \approx 0.036$ for $J_2/J_1 = 1.95$).

\begin{figure}[t]
\begin{center}
\includegraphics[width=0.51\textwidth]{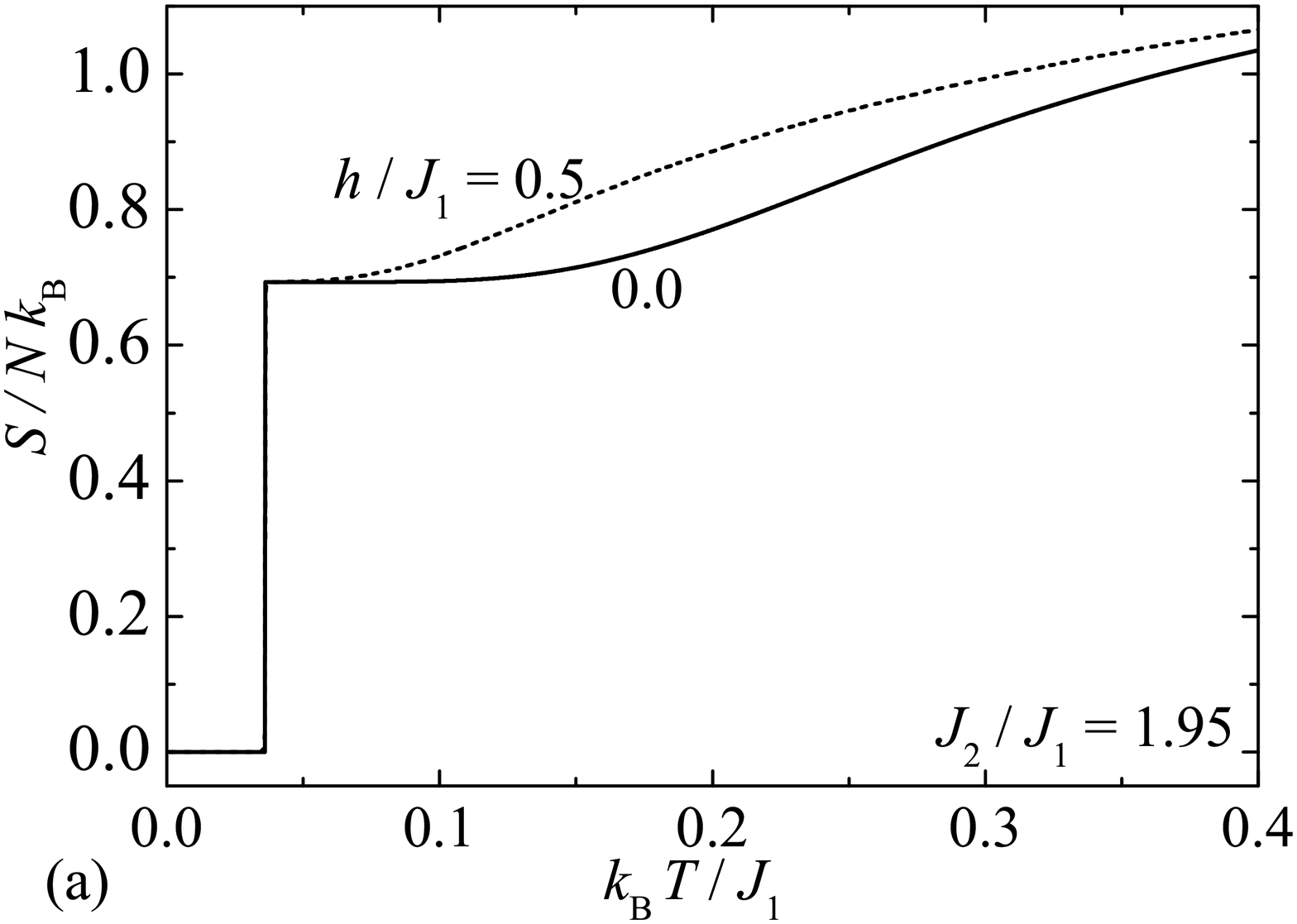}
\hspace*{-0.6cm}
\includegraphics[width=0.51\textwidth]{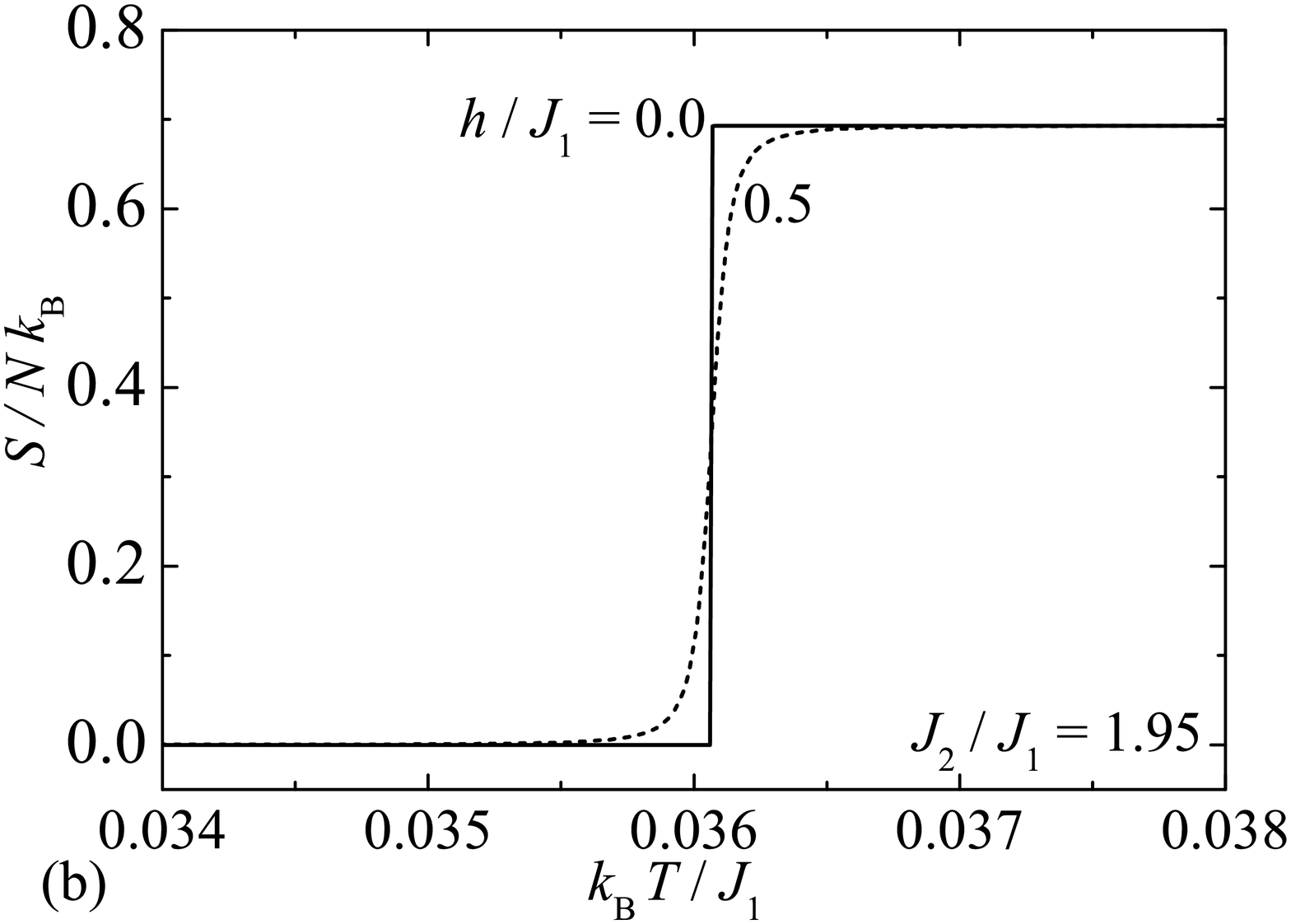}
\end{center}
\vspace{-0.7cm}
\caption{The temperature dependence of the entropy of the spin-1/2 Ising tetrahedral chain for the particular value of the interaction ratio $J_2/J_1 = 1.95$ and two different values of the magnetic field $h/J_1 = 0.0$ and $0.5$. Panel (b) depicts a detailed plot of the entropy in a vicinity of the pseudo-transition, where the entropy exhibits an abrupt but still continuous change quite reminiscent of a discontinuous jump.}
\label{tcen}
\end{figure}

To support this statement, the entropy of the spin-1/2 Ising tetrahedral chain is plotted in Fig. \ref{tcen} versus temperature for the fixed value of the interaction ratio $J_2/J_1 = 1.95$ and two different values of the magnetic field $h/J_1 = 0.0$ and $0.5$. As one can see from Fig. \ref{tcen}, the entropy of the spin-1/2 Ising tetrahedral chain indeed displays an abrupt but still continuous change at the pseudo-critical temperature $k_{\rm B} T_p/J_1 \approx 0.036$, which is successively followed by a quasi-plateau at the specific value of the entropy $S \approx N k_{\rm B} \ln 2$ in concordance with the residual entropy of the frustrated antiferromagnetic phase. It is worth noticing, moreover, that thermal dependencies of the entropy at two different magnetic fields shown in Fig. \ref{tcen}(a)-(b) indicate a gradual melting of the pseudo-transition due to a strengthening of the external magnetic field. 

\begin{figure}[t]
\begin{center}
\includegraphics[width=0.51\textwidth]{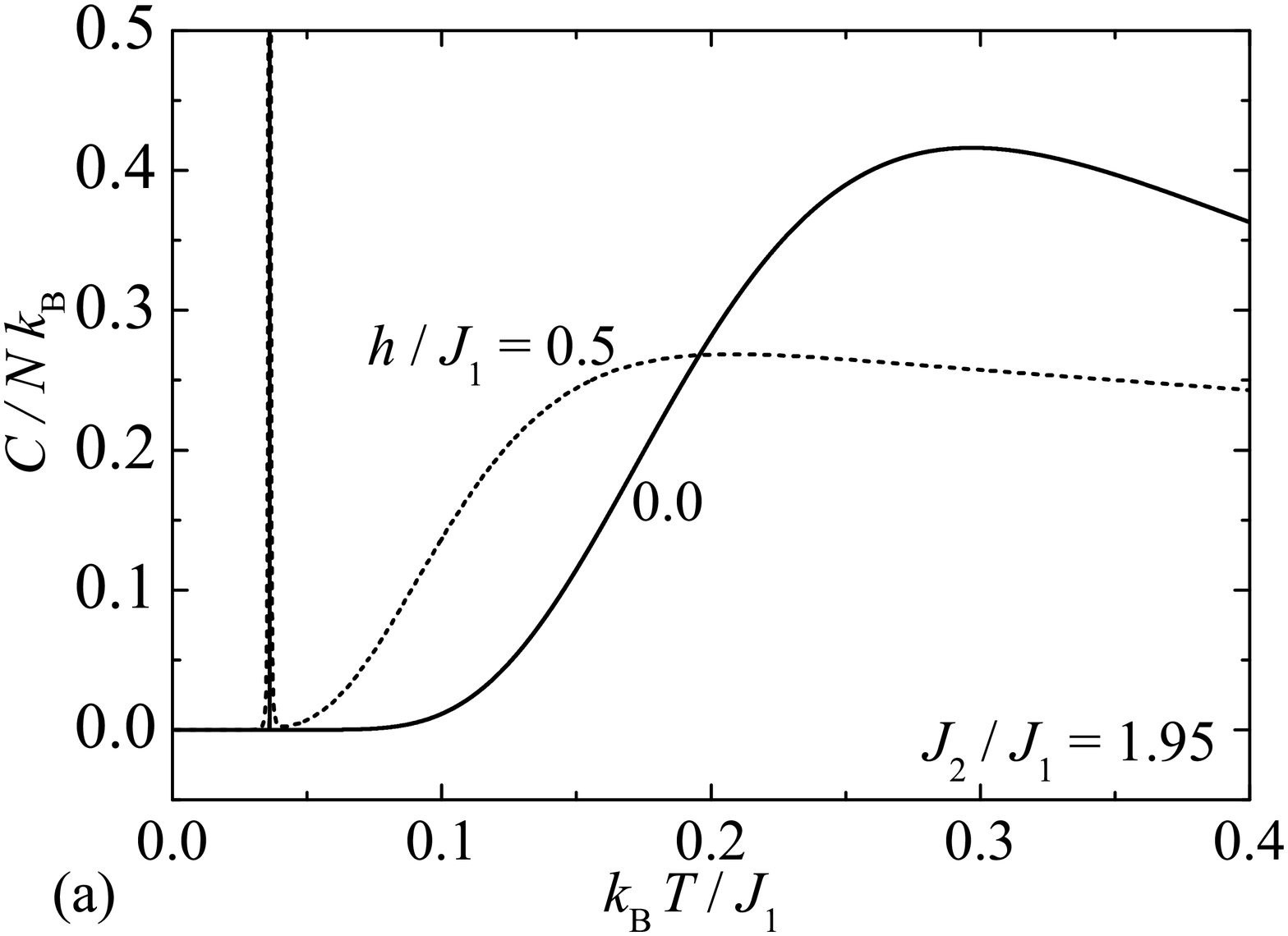}
\hspace*{-0.6cm}
\includegraphics[width=0.51\textwidth]{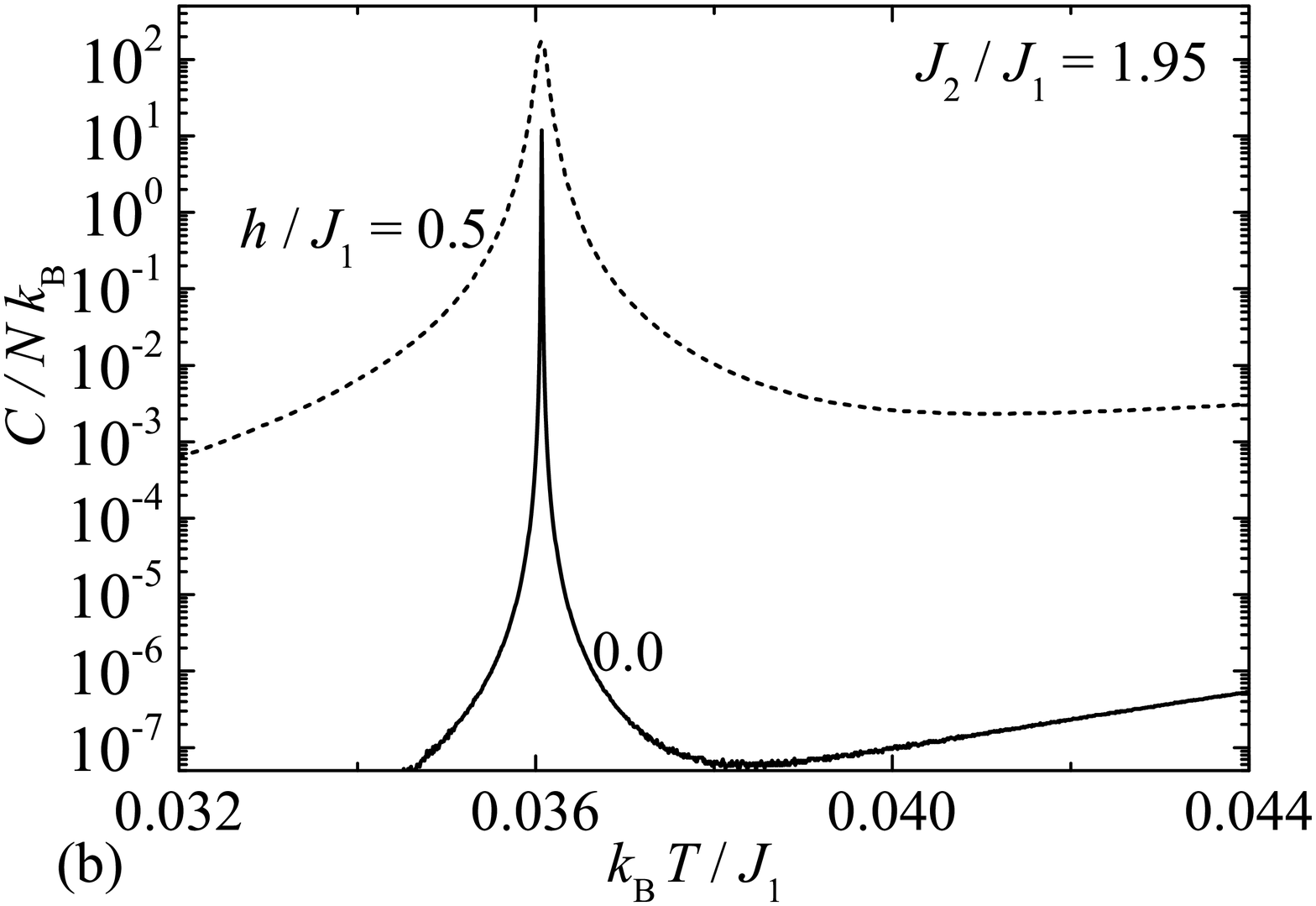}
\end{center}
\vspace{-0.7cm}
\caption{The temperature dependence of the specific heat of the spin-1/2 Ising tetrahedral chain for the particular value of the interaction ratio $J_2/J_1 = 1.95$ and two different values of the magnetic field $h/J_1 = 0.0$ and $0.5$. Panel (b) shows a semi-logarithmic plot of the specific heat in a vicinity of the pseudo-transition, where the specific heat displays a sharp finite peak quite reminiscent of a power-law divergence.}
\label{tcsh}
\end{figure}

The pseudo-transition of the spin-1/2 Ising tetrahedral chain apparently manifest itself also in anomalous temperature variations of the specific heat, which are displayed in Fig. \ref{tcsh} for one fixed value of the interaction ratio $J_2/J_1 = 1.95$ and two different values of the magnetic field $h/J_1 = 0.0$ and $0.5$. It is evident from Fig. \ref{tcsh}(a) that the specific heat shows a remarkable double-peak temperature dependence, whereas a very sharp intense peak observable at lower temperature is well separated from a round high-temperature maximum. The melting of pseudo-transition achieved upon strengthening of the magnetic field is consistent with a less sharp peak of the specific heat. Although the sizable peak of the specific heat at a pseudo-critical temperature is somewhat reminiscent of a power-law divergence, this vigorous peak is just finite and originates from intense thermal excitations from the quasi-superantiferromagnetic (qSAF) phase to the quasi-frustrated-antiferromagnetic (qFRU) phase instead of a true continuous temperature-driven phase transition. 

\begin{figure}[t]
\begin{center}
\includegraphics[width=0.51\textwidth]{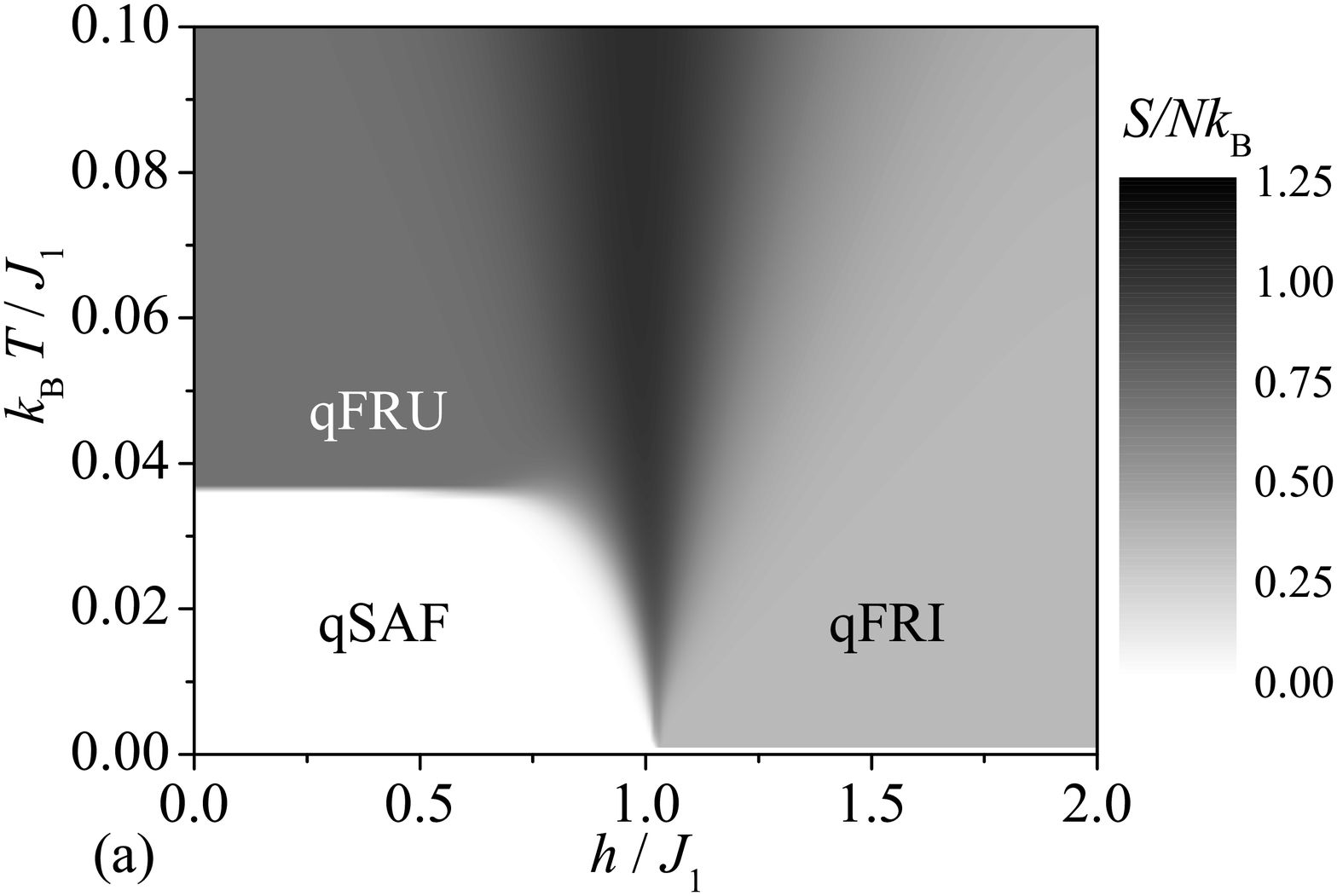}
\hspace*{-0.6cm}
\includegraphics[width=0.51\textwidth]{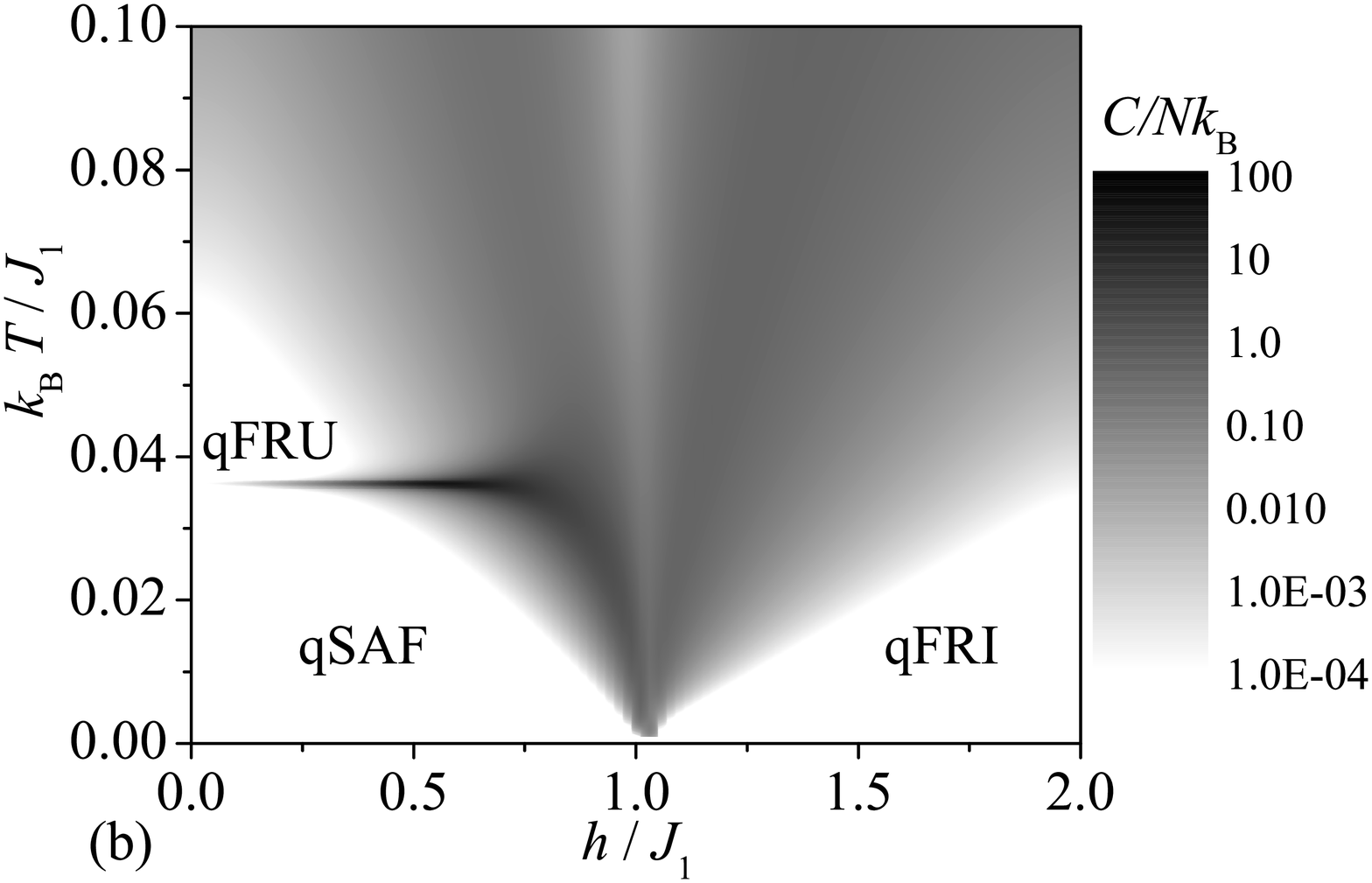}
\end{center}
\vspace{-0.7cm}
\caption{A density plot of the entropy [panel (a)] and specific heat [panel (b)] of the spin-1/2 Ising tetrahedral chain in the field-temperature ($h/J_1-k_{\rm B} T/J_1$) plane for the particular value of the interaction ratio $J_2/J_1 = 1.95$.}
\label{tc3des}
\end{figure}

A survey of theoretical results for the pseudo-critical behavior of the spin-1/2 Ising tetrahedral chain in a magnetic field will be concluded by a comprehensive analysis of the density plots of the entropy and specific heat, which are depicted in Fig. \ref{tc3des} for one representative value of the interaction ratio $J_2/J_1 = 1.95$. It is obvious from Fig. \ref{dc3dss}(a) that the density plot of entropy convincingly evidences of the pseudo-transition between the quasi-superantiferromagnetic (qSAF) phase to the quasi-frustrated-antiferromagnetic (qFRU) phase, which is accompanied with an abrupt change of the entropy at the pseudo-critical temperature $k_{\rm B} T_p/J_1 \approx 0.036$. While the pseudo-transition is relatively sharp at lower magnetic fields $h/J \lesssim 0.5$, i.e. it resembles a discontinuous temperature-driven phase transition, the pseudo-transition substantially melts at higher magnetic fields $h/J \gtrsim 0.5$ as it overlaps with a quasi-critical fan emanating from a quasi-phase boundary between the quasi-superantiferromagnetic (qSAF) phase and the quasi-frustrated-ferrimagnetic (qFRI) phase both having nonzero residual entropy. 

On the other hand, the density plot of specific heat shown in Fig. \ref{tc3des}(b) evidences the pseudo-transition between the quasi-superantiferromagnetic (qSAF) phase and the quasi-frustrated-antiferromagnetic (qFRU) phase through a vigorous increase of the specific heat spread over several orders of magnitude in a relatively tiny temperature interval near the pseudo-critical temperature $k_{\rm B} T_p/J_1 \approx 0.036$. The density plot of specific heat thus furnishes a finite-temperature phase diagram of the spin-1/2 Ising tetrahedral chain in a magnetic field, which allows to discern parameter regions for all individual quasi-phases in the most authentic manner. The quasi-superantiferromagnetic (qSAF) phase is bounded from above by the pseudo-transition with the quasi-frustrated-antiferromagnetic (qFRU) phase and from the right by a quasi-critical fan emanating from the quasi-phase boundary with the quasi-frustrated-ferrimagnetic (qFRI) phase. Furthermore, the quasi-frustrated-antiferromagnetic (qFRU) phase can be allocated within a relatively narrow (white) parameter region emergent at small enough magnetic fields $h/J \lesssim 0.5$ and temperatures $0.036 \lesssim k_{\rm B} T/J_1 \lesssim 0.065$.  

\section{Conclusion}

In this chapter we have comprehensively studied a pseudo-critical behavior of the spin-1/2 Ising diamond and tetrahedral chains by a detailed examination of basic magnetothermodynamic quantities such as the entropy, specific heat and susceptibility. Although one-dimensional lattice-statistical spin models with short-range interactions cannot exhibit a true temperature-driven phase transition we have furnished a rigorous proof that the spin-1/2 Ising diamond and tetrahedral chains in a magnetic field may display a remarkable pseudo-transition whenever the investigated spin chains are driven close enough to a ground-state phase boundary between two phases with very different entropy. Hence, the observed pseudo-criticality has been connected to intense thermal excitations from a ground state with sufficiently small entropy to a highly degenerate manifold of low-lying excited state with nonzero residual entropy 
(i.e. huge macroscopic degeneracy). It could be thus concluded that the remarkable pseudo-transitions of the spin-1/2 Ising diamond and tetrahedral chains in a magnetic field are entropy-driven phenomena and a substantial difference between degeneracies of two respective quasi-phases is an essential prerequisite for observation of a relevant pseudo-critical behavior. 

It is worthwhile to mention that the pseudo-transition of spin-1/2 Ising diamond and tetrahedral chains is clearly manifested in an anomalous behavior of basic magnetothermodynamic quantities, which mimic a temperature-driven phase transition either of a discontinuous (entropy) or continuous (specific heat, susceptibility) nature in spite of the fact that these quantities do not actually exhibit true discontinuities or power-law divergences at a pseudo-critical temperature. It has been thus demonstrated that the density plots of entropy, specific heat and susceptibility provide a useful tool for construction of finite-temperature phase diagram between the relevant quasi-phases, which is of particular interest with regard to a potential experimental verification of this peculiar phenomenon. 

It is also noteworthy that the specific heat and susceptibility follow within a relatively tiny temperature interval sufficiently close (but not too close) to a pseudo-critical temperature a power-law dependence, which is characterized by the universal values of the pseudo-critical exponents $\alpha=\alpha'=\gamma=\gamma'=3$ in agreement with the recent conjecture reported for a class of one-dimensional lattice-statistical spin models displaying a pseudo-transition \cite{roj19}. The spin-1/2 Ising diamond and tetrahedral chains in a magnetic field thus provide two valuable examples of one-dimensional lattice-statistical spin models, which exhibit a marked pseudo-critical behavior in spite of their fully classical nature. It is supposed that the similar findings could be overlooked in several frustrated one-dimensional Ising spin models, which may possibly display a pseudo-transition in a relatively tiny range of temperature and magnetic fields.

\section*{Acknowledgement}
This work was financially supported by a grant of the Ministry of Education, Science, Research and Sport of the Slovak Republic under Contract No. VEGA 1/0531/19 and by a grant of the Slovak Research and Development Agency under Contract No. APVV-16-0186.

Reviewed by: \\
\\
Taras Verkholyak, \\
Institute for Condensed Matter Physics, \\
National Academy of Sciences of Ukraine, \\ 
L'viv, Ukraine \\
\\
Lucia G\'alisov\'a, \\
Faculty of Manufacturing Technologies with the seat in Pre\v{s}ov,\\
Technical University of Ko\v{s}ice,\\
Pre\v{s}ov, Slovakia \\
\\
Olesia Krupnitska, \\
Institute for Condensed Matter Physics, \\
National Academy of Sciences of Ukraine, \\ 
L'viv, Ukraine \\
\\

\label{lastpage-01}

\end{document}